# Simulations of decomposition kinetics of Fe-Cr solid solutions during thermal aging

E. Martinez[1a], C.-C. Fu[2b], M. Levesque[2c], M. Nastar[2d] and F. Soisson[2e]

[1]MST-8, EFRC, Los Alamos National Laboratory, Los Alamos, NM 87544, USA

[2]CEA, DEN, Service de Recherches de Métallurgie Physique, 91191 Gif-sur-Yvette, France

[a]enriquem@lanl.gov, [b]chuchun.fu@cea.fr, [c]maximilen.levesque@cea.fr, [d]maylise.nastar@cea.fr, [e]frederic.soisson@cea.fr

**Keywords:** Precipitation Kinetics. Fe-Cr alloys. Monte Carlo simulations.

**Abstract.** The decomposition of Fe-Cr solid solutions during thermal aging is modeled by Atomistic Kinetic Monte Carlo (AKMC) simulations, using a rigid lattice approximation with composition dependant pair interactions that can reproduce the change of sign of the mixing energy with the alloy composition. The interactions are fitted on *ab initio* mixing energies and on the experimental phase diagram, as well as on the migration barriers in iron and chromium rich phases. Simulated kinetics is compared with 3D atom probe and neutron scattering experiments.

**Introduction**

High chromium ferritic and martensitic steels are considered as essential materials for several concepts of the next generation of fission reactors [1] as well as for future fusion reactors [2]. The addition of Cr has a positive effect on mechanical, corrosion and radiation resistance properties, so that the stability of Fe-Cr solid solutions versus the α-α' decomposition is of great technological importance. The thermodynamic and kinetic modeling of this unmixing process is a challenging problem, even without irradiation, due to the specific properties of Fe-Cr alloys which present a unique inversion of short range order [3] for the Cr contents of technological interest (9-12%). This behavior is related to the magnetic properties of these alloys, as it has been shown by recent *ab initio* studies (see e.g. Ref. [4,5]). We propose here a simple diffusion model of vacancy diffusion using composition dependent pair interactions – partly fitted on *ab initio* calculations – that can be directly used in Monte Carlo simulations to model the decomposition kinetics.

**Thermodynamic and Diffusion model**

**DFT calculations of mixing energies.** The mixing energies $\Delta E_{mix}$ of special quasi-random structures (SQS) have been computed with the PWSCF code using the PAW approximation. Results are shown on Fig. 1. As in previous DFT studies, $\Delta E_{mix}$ are negative for low Cr contents and positive for higher contents.

Several interaction models have been proposed to reproduce this energetic landscape of Fe-Cr alloys. Simple pair interaction models are not suitable, since they lead to symmetrical thermodynamic properties. A classical "chemical" cluster expansion, including many-body interactions [6], could be consider, but these additional interactions would considerably increase the time needed to compute the migration barrier. A magnetic-chemical cluster expansion [7] would be even better, since it would take into account the main physical ingredients controlling the thermodynamics of Fe-Cr alloys, but it would require to perform simulations with relaxation of the distribution of atomic species and magnetic moments, two processes that could have very different time scales: to our knowledge it has never been attempted in Kinetic Monte Carlo simulations. Previous simulations of decomposition kinetics have been proposed using non-magnetic semi-empirical potentials [8,9] fitted on DFT mixing energies. With such potential, the calculation of the migration barriers can in principle take into account atomic relaxations and long range elastic interactions. However, such calculations remain computationally demanding, and simulations of ref. [8,9] have been carried out on a perfect lattice. Moreover, it is quite difficult to develop potentials that reproduce adequate values of the key thermodynamic and kinetic properties controlling the precipitation kinetics [10].

**Thermodynamic model.** For AKMC simulations, the simplest model that can reproduce the asymmetrical mixing energies of the DFT calculations is probably a rigid lattice model with composition dependent pair interactions energies. The system energy is computed as a sum of first $\varepsilon_{ij}^{(1)}$ and second $\varepsilon_{ij}^{(2)}$ nearest interactions ($i, j$ = A, B or V). The hetero-atomic interactions $\varepsilon_{AB}^{(1)}$ and $\varepsilon_{AB}^{(2)}$ depend on the local Cr concentration, measured on the sets of first and second nearest neighbors of each pair. In the Bragg-Williams approximation, the mixing energy of an $A_{1-x}B_x$ solid solution is then:

$$E_{mix} = \Omega(x)x(1-x) \qquad (1)$$

Where $\Omega(x)$, the alloy ordering energy is :

$$\Omega(x) = \frac{z_1}{2}\left[\varepsilon_{AA}^{(1)} + \varepsilon_{BB}^{(1)} - 2\varepsilon_{AB}^{(1)}(x)\right] + \frac{z_2}{2}\left[\varepsilon_{AA}^{(2)} + \varepsilon_{BB}^{(2)} - 2\varepsilon_{AB}^{(2)}(x)\right] \qquad (2)$$

$\Omega(x)$ is assumed here to have a polynomial expression and is fitted [eq. (1)] on the mixing energies of Special Quasi-random Structures SQS (Fig. 1, where the mixing energies of various ordered structures are also shown fir comparison). The resulting phase diagram is determined by Monte Carlo calculations in the semi-grand canonical ensemble (Fig. 2): it is asymmetrical, with a solubility limit at low temperature close to the one proposed in the recent reviews by Malerba et al [11] and Xiong et al [12]. However, the critical temperature is much larger than the experimental one (~2500K): a linear dependence of the pair interactions $\varepsilon_{AB}^{(1)}(x,T)$ and $\varepsilon_{AB}^{(2)}(x,T)$ on the temperature is then introduced, which is a way to take into account (in a crude way) non-configurational entropic contributions, i.e. the magnetic and vibrational degrees of freedom of Fe-Cr alloys. By fitting these contributions on the experimental critical temperature, one gets a miscibility gap in good agreement with the experimental ones (Fig. 2).

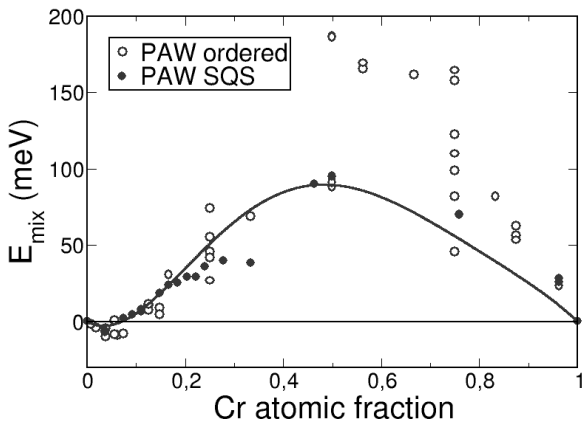 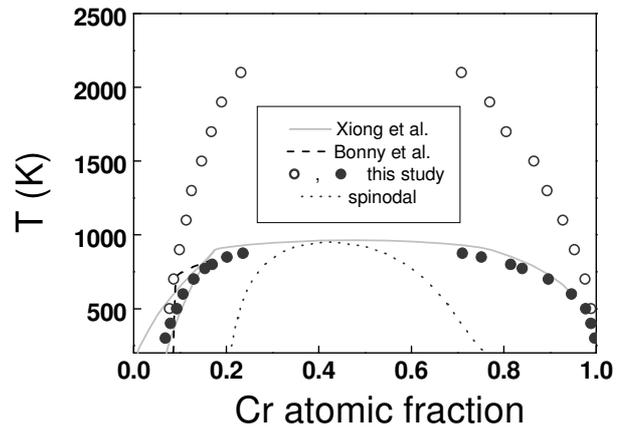

Figure 1: Mixing energies of some Special Quasi-random Structures (SQS) calculated with PWSCF-PAW and fitted by Eq. (1) with a polynomial ordering energy $\Omega(x)$.

Figure 2: The phase diagram corresponding to the ordering energy fitted on the mixing energies of Fig. 1, (○) without and (●) with non-configurational entropy contributions.

**Diffusion model.** The vacancy migration barriers are computed in a simple broken bond model, the activation energy for a given transition being simply the balance between the bonds of the bcc lattice broken during a jump and the bonds created at the saddle-point positions between the jumping atom and its neighbors [13]. In addition to the previous interatomic pair interactions $\varepsilon_{ij}^{(n)}$ (with their composition and temperature dependence), constant vacancy-atom interactions $\varepsilon_{iV}^{(n)}$ are introduced to control the vacancy formation and the binding energies. Effective saddle-point interactions $\varepsilon_{ij}^{(SP)}$ control the migration barriers. We have fitted these parameters on vacancy formation energies and on the Fe and Cr migration barriers in pure iron and pure chromium. DFT calculations of the migration barriers have been also used to compute the impurity diffusion coefficients of Cr in Fe and Fe in Cr, using the 10 frequencies Le Claire's diffusion model [14]. Finally, the attempt frequencies of Fe-V and Cr-V exchange have been fitted on the pre-exponential factors of the diffusion coefficients.

In ferromagnetic α-iron, DFT calculations give a vacancy formation energy $E_V^{for}(Fe) = 2.2$ eV and a migration barrier $E_V^{mig}(Fe) = 0.7$ eV, which correspond to a self-diffusion energy $Q_{Fe} = E_V^{mig}(Fe) + E_V^{for}(Fe) = 2.9$ eV, in good agreement with experimental results [15] and previous ab initio calculations. In pure chromium, $E_V^{for}(Cr) = 1.9$ eV and $E_V^{mig}(Cr) = 1.3$ eV have been estimated respectively in the most stable Spin Density Wave (SDW) configuration of Cr at 0 K and in the antiferromagnetic configuration. It is worth noting that the vacancy formation energy in SDW chromium is significantly lower than the one computed in the antiferromagnetic phase ($E_V^{for}(Cr) = 2.5$ eV) and closer to the experimental value [16]. The ageing temperatures are well above the Néel temperature, so it is difficult to choose the most appropriate value for modeling the diffusion in paramagnetic Cr. There are few experimental studies and they give quite different values for the self-diffusion in paramagnetic Cr [16]: direct measurement at high temperatures (>1200 K) give $Q_{Cr} \simeq 4.6$ eV and are not compatible with a monovacancy diffusion mechanism. At low temperature, positron annihilation studies suggest $Q_{Cr} \simeq 3.1$ eV, in good agreement with our estimation.

**Monte Carlo Simulations.** Monte Carlo simulations have been performed in simulations boxes of a few millions of bcc sites, with one vacancy periodic boundary conditions, using the residence time algorithm and the time rescaling procedure explained in Ref. [13], and modified to take into account to dependence of the pair interactions on the local composition. The time rescaling is performed to correct the difference between the vacancy concentration in the simulation box and the equilibrium one [13]. It must be emphasized that if some AKMC parameters are partly fitted on experimental data (pair interaction on the solubility limit and attempt frequency on the diffusion coefficient), the simulated decomposition kinetics and its time scale are not fitted on the experimental kinetics.

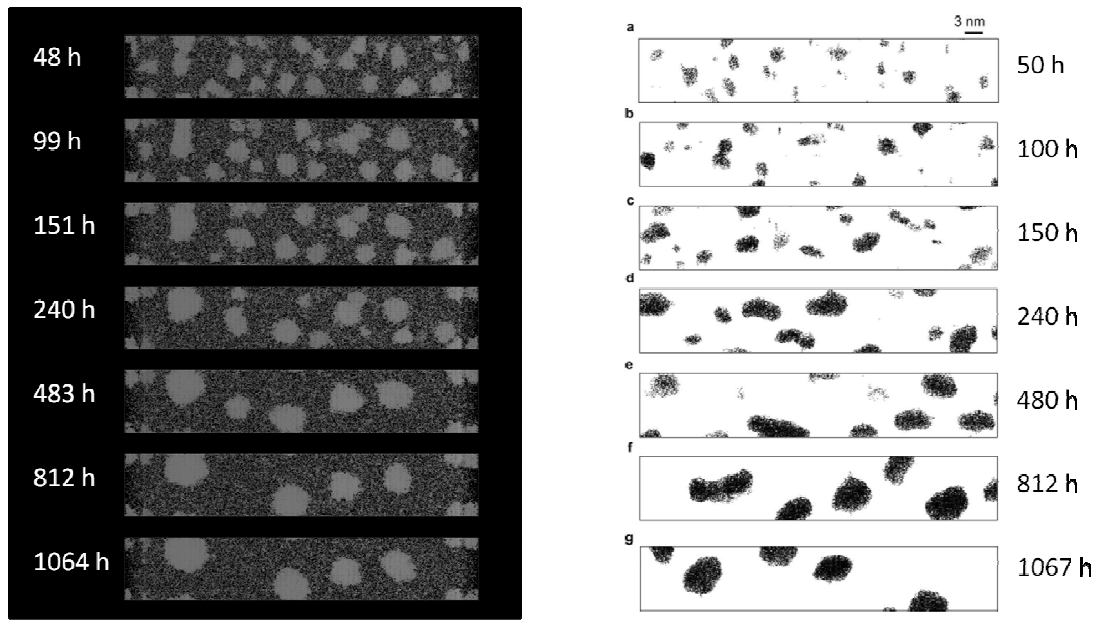

Figure 3: Decomposition kinetics in Fe-20%at.Cr at 500°C. (a) Monte Carlo simulations (64×64×352 bcc lattice sites), (b) 3D Atom Probe experiments of Novy et al. [17]

## Comparison with 3D atom probe

Novy et al [17] have recently observed the decomposition kinetics in a binary Fe-20%Cr alloy at 500°C, by 3D atom probe. AKMC simulations have been performed in the same conditions and are compared to the experimental results on Fig. 3 and 4, for similar ageing times. It can be seen that the evolution of the size and density of Cr rich clusters are in good agreement, although the average precipitate size seems a little larger in the simulations, especially before 100 h.

For ageing times longer than approximately 500 h, the composition of α and α' phases observed in the simulations are very close to those measured by 3D atom probe (Fig. 4). However, the evolution at shorter times in much more rapid in the simulation, particularly for the α' phase. The differences are important for small precipitates only (with a radius below 1.5 nm). Further analysis is needed to determine the origin of these discrepancies. They could result from a poor description, in the simulation, of the diffusion properties in the transient concentrated solid solution; or from uncertainties in the 3DAP measurements. For precipitate radius smaller than 2 nm, the overlapping of ion trajectories and the small number of atoms, may lead to significant errors [18].

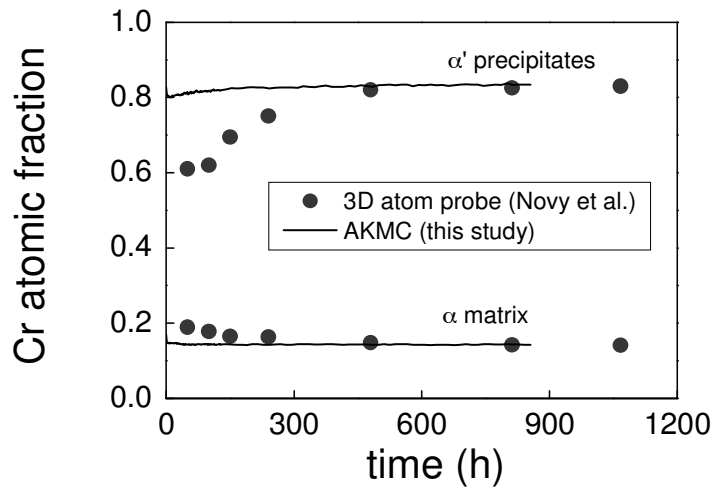

Figure 4: Evolution of the composition of α and α' phases during the decomposition of Fe-20%Cr at 500°C. Comparison between the Monte Carlo simulations (64×64×352 bcc sites) and the 3D atom probe experiments of Novy et al. [17]

## Comparison with SANS

Small Angle Neutrons scattering experiments have been used to follow the decomposition kinetics of Fe-Cr alloys at several compositions and temperatures [19,20]. The evolution of the position of the maximum of scattering intensity $q_m$ is especially interesting, because it can be directly compared to the maximum of the structure factor calculated in the simulations. At 500°C the evolution of $q_m$ has been studied for Cr contents of 20%, 35% and 50% [19]: the AKMC results are in good agreement at 20 and 35% of Cr (Fig. 5). The agreement is still reasonable at 50% (with a shift of ~ 2 on the timescale, again without fitting any parameter on the decomposition kinetics). One can notice that in the simulations as in the SANS experiments, the Cr content has only a small effect on the time evolutions of $q_m$. In Fe-20%Cr, the evolution of the microstructure observed in the simulations displays a nucleation and growth of isolated Cr-rich precipitates, while in Fe-35%Cr and Fe-50%Cr, the evolution reveals the formation of an interconnected network of α and α' phases: further analysis is needed to compare this evolution with the theories of spinodal decomposition.

At a slightly higher temperature (540°C), the simulated kinetics is much slower than the one observed by SANS experiments of Furusaka et al [20]: the evolution of $q_m$ appears to be more accelerated by a factor ~10 between 500°C and 540°C in the Monte Carlo simulations (which can be essentially explained by a similar increase of the diffusion coefficients), while the acceleration factor is ~100 according to the experiments (Fig. 5(c)). For the time being, this discrepancy is not well understood. It could be due to an error in the solubility limit at 540°C, but it seems in contradiction with the fact that the evolution of $q_m$ is not very dependent of the supersaturation at 500°C. Another explanation could be the effect of the initial configuration: we have used a random solid solution in the simulations, while some initial short range order could exist in the experiments, depending on the homogenization treatments.

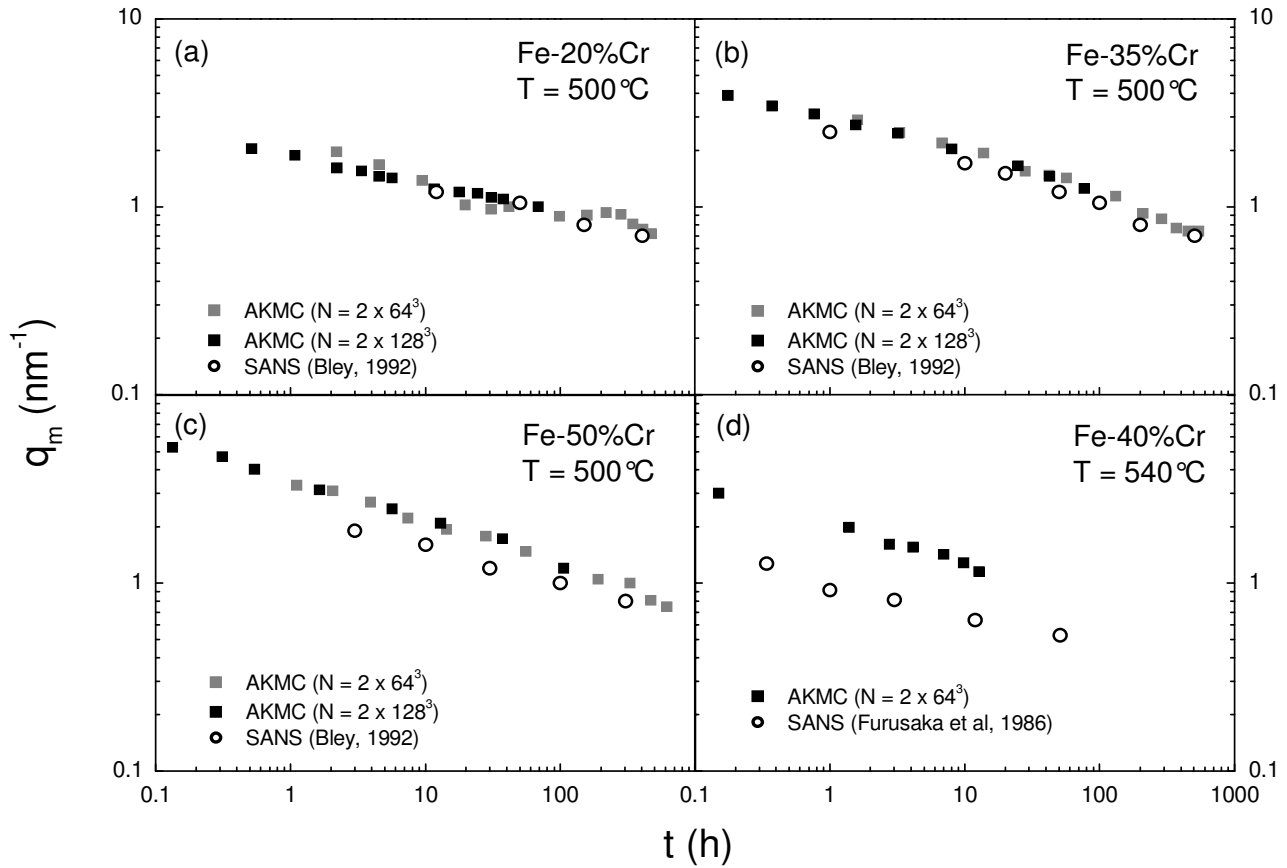

Figure 5: Evolution of the position of the peak of scattering intensity in SANS experiments of Bley [19] at 500°C for Cr contents of (a) 20%, (b) 35%, (c) 50%; and of Furusaka et al. [20] at 540°C for a Cr content of 40% (d). Comparison with AKMC simulations (with $N$ bcc lattice sites)

**Conclusion**

Preliminary Monte Carlo simulations of α-α' decomposition kinetics of iron-chromium solid solutions show that the experimental behavior can be reasonably reproduced, at a given temperature in a wide range of composition, by a simple model of composition dependent pair interaction model. Further study is still needed to understand the evolution of phase compositions and the effect of the initial short range order. One can then expect that this model will set a basis for future study of radiation effects on segregation and decomposition kinetics in these alloys. This will require the introduction in the simulation of specific irradiation events, especially the diffusion of self-interstitial atoms with their migration mechanisms.

**Aknowledgements**. This research received partial support from the Euratom FP7 (Grant No. 2007–2011), Grant Agreement No. 212175 (GetMat project). It was partly performed using HPC resources from GENCI (Grant 2009-x2009096020 and 2010-x2010096020).